\documentstyle[12pt,fleqn]{article}

\font\twlgot =eufm10 scaled \magstep1
\font\egtgot =eufm8
\font\sevgot =eufm7

\font\twlmsb =msbm10 scaled \magstep1
\font\egtmsb =msbm8
\font\sevmsb =msbm7

\newfam\gotfam
\def\pgot{\fam\gotfam\twlgot}
\textfont\gotfam\twlgot
\scriptfont\gotfam\egtgot
\scriptscriptfont\gotfam\sevgot
\def\got{\protect\pgot}

\newfam\msbfam
\textfont\msbfam\twlmsb
\scriptfont\msbfam\egtmsb
\scriptscriptfont\msbfam\sevmsb
\def\Bbb{\protect\pBbb}
\def\pBbb{\relax\ifmmode\expandafter\Bb\else\typeout{You cann't use
Bbb in text mode}\fi}
\def\Bb #1{{\fam\msbfam\relax#1}}

\newcommand{\gO}{{\got O}}
\newcommand{\gU}{{\got U}}
\newcommand{\gE}{{\got E}}
\newcommand{\gQ}{{\got Q}}

\def\thebibliography#1{\section*{References}\list
  {[\arabic{enumi}]}{\settowidth\labelwidth{#1}\leftmargin\labelwidth
    \advance\leftmargin\labelsep
    \usecounter{enumi}}
    \def\newblock{\hskip .11em plus .33em minus .07em}
    \sloppy\clubpenalty4000\widowpenalty4000
    \sfcode`\.=1000\relax}

\def\op#1{\mathop{\fam0 #1}\limits}

\newcommand{\id}{{\rm Id\,}}

\newcommand{\di}{{\rm dim\,}}

\newcommand{\Ker}{{\rm Ker\,}}
\newcommand{\nm}[1]{\mid {#1}\mid}

\newcommand{\beq}{\begin{equation}}
\newcommand{\eeq}{\end{equation}}
\newcommand{\ben}{\begin{eqnarray}}
\newcommand{\een}{\end{eqnarray}}
\newcommand{\be}{\begin{eqnarray*}}
\newcommand{\ee}{\end{eqnarray*}}
\newcommand{\bea}{\begin{eqalph}}
\newcommand{\eea}{\end{eqalph}}

\newcommand{\cL}{{\cal L}}
\newcommand{\cO}{{\cal O}}

\newcommand{\cQ}{{\cal Q}}
\newcommand{\cE}{{\cal E}}

\newcommand{\bL}{{\bf L}}

\newcommand{\al}{\alpha}

\newcommand{\dl}{\delta}
\newcommand{\la}{\lambda}
\newcommand{\La}{\Lambda}
\newcommand{\f}{\phi}
\newcommand{\om}{\omega}

\newcommand{\m}{\mu}

\newcommand{\g}{\gamma}
\newcommand{\G}{\Gamma}
\newcommand{\th}{\theta}

\newcommand{\si}{\sigma}
\newcommand{\Si}{\Sigma}
\newcommand{\w}{\wedge}

\newcommand{\ol}{\overline}
\newcommand{\dr}{\partial}
\newcommand{\ar}{\op\longrightarrow}
\newcommand{\ot}{\otimes}
\newcommand{\ap}{\approx}

\newcounter{eqalph}
\newcounter{equationa}
\newcounter{theorem}
\newcounter{remark}
\newcounter{proposition}
\newcounter{lemma}
\newcounter{corollary}
\newcounter{definition}

\newenvironment{eqalph}{\stepcounter{equation}
\setcounter{equationa}{\value{equation}}
\setcounter{equation}{0}

\begin{eqnarray}}{\end{eqnarray}\setcounter{equation}{\value{equationa}}}

\def\theremark{\arabic{remark}}

\def\thedefinition{\arabic{definition}}

\newenvironment{proof}{\noindent 
{\bf Proof.}}{\hfill $\Box$ \medskip}
\newenvironment{rem}{\refstepcounter{remark}\medskip\noindent{\bf
Remark \theremark.} }{\medskip}
\newenvironment{theo}{\refstepcounter{definition} 
\medskip\noindent{\bf Theorem \thedefinition.}\it }{\medskip}
\newenvironment{prop}{\refstepcounter{definition} 
\medskip\noindent{\bf Proposition \thedefinition.}\it }{\medskip}
\newenvironment{lem}{\refstepcounter{definition} 
\medskip\noindent{\bf Lemma \thedefinition.}\it }{\medskip}
\newenvironment{cor}{\refstepcounter{definition} 
\medskip\noindent{\bf Corollary \thedefinition.}\it }{\medskip}

\newcommand{\mar}[1]{}
\begin{document}
\hbox{}

\thispagestyle{empty}

{\parindent=0pt

{\large\bf Cohomology of the variational complex}
\bigskip

{\sc Giovanni Giachetta$\dagger$, Luigi Mangiarotti$\dagger$ and Gennadi 
Sardanashvily$\ddagger$}
\medskip

\begin{small}
$\dagger$ Department of Mathematics and Physics, University of Camerino, 62032
Camerino (MC), Italy \\
$\ddagger$ Department of Theoretical Physics, Physics Faculty, Moscow State
University, 117234 Moscow, Russia
\medskip

E-mail: giachetta@campus.unicam.it, mangiaro@camserv.unicam.it and
sard@grav.phys.msu.su

\bigskip\bigskip

{\bf Abstract.}
Cohomology of the variational bicomplex 
in the calculus of
variations in classical field theory are computed in the class of
exterior forms of finite jet order. This
provides a solution of the global inverse problem of the finite order calculus
of variations.
\bigskip


\end{small}

}

\bigskip

\section{Introduction}

Let $Y\to X$ be a smooth fibre bundle of a field model. We study cohomology of
the  variational bicomplex of exterior forms on the infinite
order jet space 
$J^\infty Y$ of $Y\to X$. The exterior differential on $J^\infty Y$ splits
into the sum of the vertical differential $d_V$ and the horizontal
differential $d_H$. These differentials, together with the variational
operator $\dl$, constitute the variational bicomplex of exterior forms on
$J^\infty Y$.

Note that the two differential algebras of exterior
forms $\cO^*_\infty$ and $\cQ^*_\infty$ are usually considered on
$J^\infty Y$.  The $\cO^*_\infty$ consists of all exterior forms on 
finite order jet manifolds modulo the pull-back identification.
Lagrangian field theory is phrased in terms of $\cO^*_\infty$. 
Its cohomology,  except de Rham cohomology and a particular
result of
\cite{vin} on $\dl$-cohomology, remains unknown. The
$\cQ^*_\infty$ is the structure algebra  of the sheaf of germs of exterior
forms on finite order jet manifolds. For short, one can say that it
consists of exterior forms of locally finite jet order. The $d_H$- and
$\dl$-cohomology of
$\cQ^*_\infty$ has been investigated in \cite{ander80,tak2}. 
Due to Lemma \ref{lmp03} below, 
we simplify this investigation
and complete it by the study of $d_V$-cohomology of $\cQ^*_\infty$. We
prove that the differential algebra 
$\cO^*_\infty$ has the same $d_H$- and $\dl$-cohomology as $\cQ^*_\infty$
(see Theorem \ref{am11} below). This provides a solution of the global
inverse problem of the calculus of variations in the class of finite
order Lagrangians. The main point for applications is that
the obstruction to the exactness of the calculus of variations is given by
closed forms on the fibre bundle
$Y$, and is of first order.

\section{The differential calculus on $J^\infty Y$}

Smooth manifolds throughout are assumed to be
real, finite-dimensional, Hausdorff,
paracompact, and connected. Put further dim$X=n\geq 1$.

Jet spaces provide the standard framework in theory of
non-linear differential equations and the calculus of variations
\cite{gol,book,pom,vin}. Recall that the $r$-order jet
space
$J^rY$ consists of sections of $Y\to X$ identified by $r+1$ terms of their
Taylor series.  The key point is that 
$J^rY$ is a smooth manifold. It is coordinated by $(x^\la,y^i,y^i_\La)$, where
$(x^\la, y^i)$ are bundle coordinates on $Y\to X$ and
$\La=(\la_k\ldots\la_1)$, $|\La|=k\leq r$, denotes a symmetric multi-index.  
The infinite order jet space $J^\infty Y$ is defined as a projective limit
$(J^\infty Y,\pi^\infty_r)$ of the inverse system
\mar{t1}\beq
X\op\longleftarrow^\pi Y\op\longleftarrow^{\pi^1_0}\cdots \longleftarrow
J^{r-1}Y \op\longleftarrow^{\pi^r_{r-1}} J^rY\longleftarrow\cdots \label{t1}
\eeq
of finite order jet manifolds $J^rY$ of $Y\to X$, where $\pi^r_{r-1}$ are
affine bundles. The surjections
\mar{5.74}\beq
\pi^\infty_r:J^\infty Y\to J^rY \label{5.74}
\eeq
obey the composition condition $\pi^\infty_i=\pi^j_i\circ\pi^\infty_j$, 
$\forall j>i$.
The set $J^\infty Y$ is provided with the coarsest
topology such that all surjections (\ref{5.74})
are continuous. The base of open sets of this topology 
consists of the inverse images of open subsets of finite order jet
manifolds  under the surjections (\ref{5.74}), which thus are open
maps. With this topology,
$J^\infty Y$ is a paracompact Fr\'echet (but not Banach) manifold
modelled on a locally convex vector space of
formal series $\{x^\la,y^i, y^i_\la,\ldots\}$ \cite{abb,tak2}.
Bearing in mind the well-known Borel theorem, one can say that 
$J^\infty Y$ consists of equivalence classes of sections of $Y\to X$ identified
by their Taylor  series at points $x\in X$.
A bundle coordinate atlas
$\{U_Y,(x^\la,y^i)\}$ of $Y\to X$ yields the manifold
coordinate atlas
\be
\{(\pi^\infty_0)^{-1}(U_Y), (x^\la, y^i_\La)\}, \qquad 0\leq|\La|,
\ee
 of $J^\infty
Y$, together with the transition functions  
\mar{55.21}\beq
{y'}^i_{\la+\La}=\frac{\dr x^\m}{\dr x'^\la}d_\m y'^i_\La, \label{55.21}
\eeq
where $\la+\La$ is the multi-index $(\la\la_k\ldots\la_1)$ and
$d_\la$ are the total derivatives 
\be
d_\la = \dr_\la + \op\sum_{|\La|\geq 0} y^i_{\la+\La}\dr_i^\La.
\ee
Moreover, $Y$ is a strong deformation retract of $J^\infty Y$ 
(see Appendix A for an explicit form of a homotopy map)

Since $J^\infty Y$ is not a Banach manifold, the familiar
geometric definition of differential objects on $J^\infty Y$ is not
appropriate (see, e.g.,
\cite{abb,tak1}). One uses the fact that $J^\infty Y$ is a projective limit of
the inverse system of manifolds (\ref{t1}). Given this inverse system, we
have the direct system 
\mar{5.7}\beq
\cO^*(X)\op\longrightarrow^{\pi^*} \cO^*_0 
\op\longrightarrow^{\pi^1_0{}^*} \cO_1^*
\op\longrightarrow^{\pi^2_1{}^*} \cdots \op\longrightarrow^{\pi^r_{r-1}{}^*}
 \cO_r^* \longrightarrow\cdots \label{5.7}
\eeq
of differential algebras $\cO^*_r$ of exterior forms on finite order jet
manifolds, where $\pi^r_{r-1}{}^*$ are pull-back monomorphisms.
This direct system
admits a direct limit
$(\cO^*_\infty,\pi^{\infty *}_r)$ in the category of $\Bbb R$-modules.
It consists of exterior forms on 
finite order jet manifolds modulo the pull-back identification, together
with the $\Bbb R$-module monomorphisms
\be
\pi^{\infty*}_k:\cO^*_k\to \cO^*_\infty 
\ee
which obey the composition condition 
$\pi^{\infty*}_i= \pi^{\infty*}_j\circ\pi^{j*}_i$, $\forall j>i$. 
Operations of the exterior product $\w$ and the exterior differentiation
$d$ of exterior forms on finite order jet manifolds
commute with the pull-back maps $\pi^r_{r-1}{}^*$ and, thus, constitute the
direct systems of the order-preserving endomorphisms of the direct system
(\ref{5.7}). These direct systems have the direct limits  which make 
$\cO_\infty^*$ a graded differential algebra. 
The $\cO^*_\infty$ is a
differential calculus over the $\Bbb R$-ring $\cO^0_\infty$ of continuous real
functions on
$J^\infty Y$ which are the pull-back of smooth real functions on
finite order jet manifolds by surjections (\ref{5.74}). 
Passing to the direct limit of de Rham complexes on finite order
jet manifolds, de Rham cohomology of the differential algebra
$\cO^*_\infty$ has only been found \cite{ander0,bau}. This
coincides with  de Rham cohomology of the fibre bundle $Y$ (see Section
4).  However, this is not a way of studying other cohomology of
the graded differential algebra $\cO^*_\infty$.  

To solve this problem,
let us enlarge $\cO^0_\infty$ to the $\Bbb
R$-ring
$\cQ^0_\infty$ of continuous real functions on $J^\infty Y$ such that, given
$f\in
\cQ^0_\infty$ and any point $q\in J^\infty Y$, there exists a neighborhood
of $q$ where $f$ coincides with the pull-back of a smooth function on some
finite order jet manifold.  The reason lies in 
the fact that the paracompact space
$J^\infty Y$ admits a partition of unity by elements of the ring
$\cQ^0_\infty$ \cite{tak2}. Therefore, sheaves of
$\cQ^0_\infty$-modules on
$J^\infty Y$ are fine and, consequently, acyclic. Then, the 
abstract de Rham theorem on cohomology of a sheaf resolution can be called
into play. 

\begin{rem}
Throughout, we follow the terminology of
\cite{hir} where by a sheaf $S$ over a topological space $Z$ is meant a sheaf
bundle $S\to Z$. Accordingly, $\G(S)$ denotes the canonical presheaf of
sections of the sheaf $S$, and 
$\G(Z,S)$ is the group of global sections of $S$. All sheaves below are
ringed spaces, but we omit this terminology if there is no danger of
confusion.
\end{rem}

Let us define a differential calculus over the ring $\cQ^0_\infty$.
Let $\gO^*_r$ be a sheaf
of germs of exterior forms on the $r$-order jet manifold $J^rY$ and 
$\G(\gO^*_r)$ its canonical presheaf.  There is the direct system of canonical
presheaves
\be
\G(\gO^*_X)\op\longrightarrow^{\pi^*} \G(\gO^*_0) 
\op\longrightarrow^{\pi^1_0{}^*} \G(\gO_1^*)
\op\longrightarrow^{\pi^2_1{}^*} \cdots \op\longrightarrow^{\pi^r_{r-1}{}^*}
 \G(\gO_r^*) \longrightarrow\cdots, 
\ee
where $\pi^r_{r-1}{}^*$ are pull-back monomorphisms
with respect to open surjections 
$\pi^r_{r-1}$. 
Its direct
limit $\gO^*_\infty$ 
is a presheaf of graded differential
$\Bbb R$-algebras on
$J^\infty Y$. 
 The germs of elements of the presheaf
$\gO^*_\infty$ constitute a sheaf
$\gQ^*_\infty$ on $J^\infty Y$. 
It means that, given a section
$\f\in\G(U,\gQ^*_\infty)$ of $\gQ^*_\infty$ over an open subset $U\subset
J^\infty Y$ and any point
$q\in U$, there exists a neighbourhood $U_q\subset U$ of $q$ such that
$\f|_{U_q}$ is the pull-back of a local exterior form on some finite order jet
manifold. 
However, $\gO^*_\infty$ does not coincide with the canonical
presheaf $\G(\gQ^*_\infty)$ the sheaf $\gQ^*_\infty$.
The structure algebra 
$\cQ^*_\infty=\G(J^\infty Y,\gQ^*_\infty)$ of
the sheaf $\gQ^*_\infty$  
is a desired differential calculus over the $\Bbb
R$-ring $\cQ^*_\infty$. 
There are obvious $\Bbb R$-algebra monomorphisms 
\be
\cO^*_\infty \to\cQ^*_\infty,  \qquad  \gO^*_\infty
\to\G(\gQ^*_\infty). 
\ee

For short, we
agree to call elements of $\cQ^*_\infty$ the
exterior forms on
$J^\infty Y$.  Restricted to a
coordinate chart
$(\pi^\infty_0)^{-1}(U_Y)$ of $J^\infty Y$, they
can be written in a coordinate form, where horizontal forms
$\{dx^\la\}$ and contact 1-forms
$\{\th^i_\La=dy^i_\La -y^i_{\la+\La}dx^\la\}$ constitute the set of
generators of the differential calculus
$\cQ^*_\infty$. 
There is the canonical splitting
\be
\cQ^*_\infty =\op\oplus_{k,s}\cQ^{k,s}_\infty, \qquad 0\leq k, \qquad
0\leq s\leq n,
\ee
of $\cQ^*_\infty$ into $\cQ^0_\infty$-modules $\cQ^{k,s}_\infty$
of $k$-contact and $s$-horizontal forms, together with the corresponding
projections
\be
h_k:\cQ^*_\infty\to \cQ^{k,*}_\infty, \quad 0\leq k, \qquad
h^s:\cQ^*_\infty\to \cQ^{*,s}_\infty, \quad 0\leq s
\leq n.
\ee 
Accordingly, the
exterior differential on $\cQ_\infty^*$ is
decomposed into the sum $d=d_H+d_V$ of horizontal and vertical
differentials such that
\be
&& d_H\circ h_k=h_k\circ d\circ h_k, \qquad d_H(\f)=
dx^\la\w d_\la(\f), \qquad \f\in\cQ^*_\infty,\\ 
&& d_V \circ h^s=h^s\circ d\circ h^s, \qquad
d_V(\f)=\th^i_\La \w \dr_\La^i\f.
\ee
They are nilpotent, i.e.,
\be
d_H\circ d_H=0, \qquad d_V\circ d_V=0, \qquad d_V\circ d_H
+d_H\circ d_V=0. 
\ee

\begin{rem}
It should be emphasized that, in the class of exterior forms of locally
finite order, all local operators are well-defined because these forms depends
locally on a finite number of variables and all sums over these variables
converge.
\end{rem} 

\begin{rem}
Traditionally, one attempts to 
introduce the differential algebra $\cQ^*_\infty$ of locally pull-back forms 
on $J^\infty Y$ in a standard geometric way
\cite{abb,bau,tak1,tak2}. The difficulty lies in the
geometric interpretation of derivations of the $\Bbb R$-ring $\cQ^0_\infty$ 
as vector fields on the Fr\'echet manifold
$J^\infty Y$.
\end{rem}

\section{The variational bicomplex}

Being nilpotent, the
differentials $d_V$ and $d_H$ provide the natural bicomplex
$\{\gQ^{k,m}_\infty\}$ of  the sheaf
$\gQ^*_\infty$ on $J^\infty Y$. To complete it to the
variational bicomplex, one considers the projection $\Bbb R$-module
endomorphism 
\be
&& \tau=\op\sum_{k>0} \frac1k\ol\tau\circ h_k\circ h^n, \\ 
&&\ol\tau(\f)
=(-1)^{\nm\La}\th^i\w [d_\La(\dr^\La_i\rfloor\f)], \qquad 0\leq\nm\La,
\qquad \f\in \G(\gQ^{>0,n}_\infty),
\ee
of $\gQ^*_\infty$ such that
\be
\tau\circ d_H=0, \qquad  \tau\circ d\circ \tau -\tau\circ d=0.
\ee
Introduced on elements of the presheaf $\gO^*_\infty$ 
(see, e.g., \cite{bau,book,tul}), this endomorphism is induced on the
sheaf $\gQ^*_\infty$ and its structure algebra
$\cQ^*_\infty$. Put
\be
\gE_k=\tau(\gQ^{k,n}_\infty), \qquad E_k=\tau(\cQ^{k,n}_\infty), \qquad k>0.
\ee
Since
$\tau$ is a projection operator, we have isomorphisms 
\be
\G(\gE_k)=\tau(\G(\gQ^{k,n}_\infty)), \qquad E_k=\G(J^\infty Y,\gE_k).
\ee
The variational operator on $\gQ^{*,n}_\infty$ is defined as the
morphism $\dl=\tau\circ d$. 
It is nilpotent, and obeys the relation 
\mar{am13}\beq
\dl\circ\tau-\tau\circ d=0. \label{am13}
\eeq

Let $\Bbb R$ and  $\gO^*_X$ denote the constant sheaf
on
$J^\infty Y$ and the sheaf of exterior forms on $X$, respectively. The
operators $d_V$,
$d_H$,
$\tau$ and $\dl$ give the following variational bicomplex of
sheaves of exterior forms on $J^\infty Y$:
\mar{7}\beq
\begin{array}{ccccrlcrlccrlccrlcrl}
& & & & _{d_V} & \put(0,-7){\vector(0,1){14}} & & _{d_V} &
\put(0,-7){\vector(0,1){14}} & &  & _{d_V} &
\put(0,-7){\vector(0,1){14}} & & &  _{d_V} &
\put(0,-7){\vector(0,1){14}}& & _{-\dl} & \put(0,-7){\vector(0,1){14}} \\ 
 &  & 0 & \to & &\gQ^{k,0}_\infty &\ar^{d_H} & & \gQ^{k,1}_\infty &
\ar^{d_H} &\cdots  & & \gQ^{k,m}_\infty &\ar^{d_H} &\cdots & &
\gQ^{k,n}_\infty &\ar^\tau &  & \gE_k\to  0\\  
 & &  &  & & \vdots & & & \vdots  & & & 
&\vdots  & & & &
\vdots & &   & \vdots \\ 
& & & & _{d_V} & \put(0,-7){\vector(0,1){14}} & & _{d_V} &
\put(0,-7){\vector(0,1){14}} & &  & _{d_V} &
\put(0,-7){\vector(0,1){14}} & & &  _{d_V} &
\put(0,-7){\vector(0,1){14}}& & _{-\dl} & \put(0,-7){\vector(0,1){14}} \\ 
 &  & 0 & \to & &\gQ^{1,0}_\infty &\ar^{d_H} & & \gQ^{1,1}_\infty &
\ar^{d_H} &\cdots  & & \gQ^{1,m}_\infty &\ar^{d_H} &\cdots & &
\gQ^{1,n}_\infty &\ar^\tau &  & \gE_1\to  0\\  
& & & & _{d_V} &\put(0,-7){\vector(0,1){14}} & & _{d_V} &
\put(0,-7){\vector(0,1){14}} & & &  _{d_V}
 & \put(0,-7){\vector(0,1){14}} & &  & _{d_V} & \put(0,-7){\vector(0,1){14}}
 & & _{-\dl} & \put(0,-7){\vector(0,1){14}} \\
0 & \to & \Bbb R & \to & & \gQ^0_\infty &\ar^{d_H} & & \gQ^{0,1}_\infty &
\ar^{d_H} &\cdots  & &
\gQ^{0,m}_\infty & \ar^{d_H} & \cdots & &
\gQ^{0,n}_\infty & \equiv &  & \gQ^{0,n}_\infty \\
& & & & _{\pi^{\infty*}}& \put(0,-7){\vector(0,1){14}} & & _{\pi^{\infty*}} &
\put(0,-7){\vector(0,1){14}} & & &  _{\pi^{\infty*}}
 & \put(0,-7){\vector(0,1){14}} & &  & _{\pi^{\infty*}} &
\put(0,-7){\vector(0,1){14}} & &  & \\
0 & \to & \Bbb R & \to & & \gO^0_X &\ar^d & & \gO^1_X &
\ar^d &\cdots  & &
\gO^m_X & \ar^d & \cdots & &
\gO^n_X & \ar^d & 0 &  \\
& & & & &\put(0,-5){\vector(0,1){10}} & & &
\put(0,-5){\vector(0,1){10}} & & & 
 & \put(0,-5){\vector(0,1){10}} & & &   &
\put(0,-5){\vector(0,1){10}} & &  & \\
& & & & &0 & &  & 0 & & & & 0 & & & & 0 & &  & 
\end{array}
\label{7}
\eeq
The second row and the last column of this bicomplex form the 
variational complex
\mar{tams1}\beq
0\to\Bbb R\to \gQ^0_\infty \ar^{d_H}\gQ^{0,1}_\infty\ar^{d_H}\cdots  
\op\longrightarrow^{d_H} 
\gQ^{0,n}_\infty  \op\longrightarrow^\dl \gE_1 
\op\longrightarrow^\dl 
\gE_2 \longrightarrow \cdots\, . \label{tams1}
\eeq
The corresponding variational bicomplexes $\{\cQ^*_\infty,E_k\}$ and
$\{\cO^*_\infty, \ol E_k\}$ of the differential calculus $\cQ^*_\infty$ and 
$\cO^*_\infty$ take place.

There are the well-known statements summarized usually as
the algebraic Poincar\'e lemma (see, e.g., \cite{olver,tul}). 

\begin{lem} \label{am12} \mar{am12}
If $Y$ is a contractible fibre bundle $\Bbb R^{n+p}\to\Bbb R^n$, the
variational bicomplex $\{\cO^*_\infty, \ol E_k\}$ of the graded differential
algebra $\cO^*_\infty$ is exact.
\end{lem}

It follows that the variational
bicomplex of sheaves (\ref{7}) is exact for any smooth fibre bundle $Y\to X$.
Moreover, all sheaves
$\gQ^{k,m}$ in this bicomplex are fine, and so are the sheaves $\gE_k$ in
accordance with the following lemma.

\begin{lem} \label{lmp03} \mar{lmp03}
Sheaves $\gE_k$, $k>0$, are fine.
\end{lem}

\begin{proof}
Though $\Bbb R$-modules $E_{k>1}$ fail to be
$\cQ^0_\infty$-modules \cite{tul}, one can use the fact that the sheaves
$\gE_{k>0}$ are projections $\tau(\gQ^{k,n}_\infty)$ of sheaves of
$\cQ^0_\infty$-modules. Let $\gU =\{U_i\}_{i\in I}$
be a 
locally finite open covering  of
$J^\infty Y$ and $\{f_i\in\cQ^0_\infty\}$ the associated partition of unity. 
For any open subset $U\subset J^\infty Y$ and any section
$\varphi$ of 
the sheaf $\gQ^{k,n}_\infty$ over $U$, let us put
$h_i(\varphi)=f_i\varphi$. Then, $\{h_i\}$ provide a family of endomorphisms
of the sheaf $\gQ^{k,n}_\infty$, required for $\gQ^{k,n}_\infty$ to be fine.
Endomorphisms $h_i$ of $\gQ^{k,n}_\infty$ also yield the $\Bbb R$-module
endomorphisms 
\be
\ol h_i=\tau\circ h_i: \gE_k\ar^{\rm in} \gQ^{k,n}_\infty \ar^{h_i}
\gQ^{k,n}_\infty \ar^\tau \gE_k
\ee
of the sheaves $\gE_k$.
They possess the properties
required for $\gE_k$ to be a fine sheaf. Indeed, for each $i\in I$, there is a
closed set ${\rm supp}\,f_i\subset U_i$ such that $\ol h_i$ is zero outside
this set, while the sum $\op\sum_{i\in I}\ol h_i$ is the identity morphism.
\end{proof}

This Lemma simplify essentially our cohomology computation of the
variational bicomplex in comparison with that in \cite{ander80,tak2}.  
Since all sheaves except $\Bbb R$ and $\pi^{\infty*}\gO^*_X$ in the bicomplex
(\ref{7}) are fine, the abstract de Rham theorem 
(\cite{hir}, Theorem 2.12.1) can be applied to columns
and rows of this bicomplex in a straightforward way. We will quote the
following variant of this theorem (see Appendix B for its proof).

\begin{theo} \label{+132} 
Let 
\mar{+131'}\beq
0\to S\ar^h S_0\ar^{h^0} S_1\ar^{h^1}\cdots\ar^{h^{p-1}} S_p\ar^{h^p}
S_{p+1}, \qquad p>1, \label{+131'}
\eeq
be an exact sequence of sheaves on a paracompact topological space $Z$, where
the sheaves $S_p$ and $S_{p+1}$ are not necessarily acyclic, and let 
\beq
0\to \G(Z,S)\ar^{h_*} \G(Z,S_0)\ar^{h^0_*}
\G(Z,S_1)\ar^{h^1_*}\cdots\ar^{h^{p-1}_*} \G(Z,S_p)\ar^{h^p_*}
\G(Z,S_{p+1}) \label{+130}
\eeq
be the corresponding cochain complex
of structure groups of these sheaves.
The $q$-cohomology groups of the
cochain complex (\ref{+130}) for $0\leq q\leq p$ are
isomorphic to the cohomology groups $H^q(Z,S)$ of $Z$ with coefficients in the
sheaf $S$. 
\end{theo}

\section{De Rham cohomology of $J^\infty Y$}

Let us start from de Rham cohomology of the graded differential algebra
$\cO^*_\infty$.

\begin{prop} \label{a1} \mar{a1}
There is an isomorphism
\be
H^*(\cO^*_\infty)= H^*(Y) 
\ee
between de Rham cohomology $H^*(\cO^*_\infty)$ of  $\cO^*_\infty$  and de Rham
cohomology 
$H^*(Y)$ of the fibre bundle
$Y$. 
\end{prop}

\begin{proof}
The proof is based on the fact that the de Rham complex
\mar{5.13}\beq
0\to \Bbb R\to
\cO^0_\infty\op\longrightarrow^d\cO^1_\infty\op\longrightarrow^d
\cdots
\label{5.13}
\eeq
of $\cO^*_\infty$ is the direct limit of de Rham
complexes of exterior forms on finite order jet manifolds. Since the exterior
differential $d$ commutes with the pull-back maps $\pi^r_{r-1}{}^*$, these
complexes form a direct system. Then, in accordance with 
the well-known theorem \cite{dold},  
the cohomology groups $H^*(\cO^*_\infty)$ of the  de Rham
complex (\ref{5.13}) are isomorphic to the direct limit  of the direct system 
\be
0\to H^*(X) \ar^{\pi^*} H^*(Y) \ar^{\pi^{1*}_0} H^*(J^1Y)\ar \cdots
\ee
of de Rham cohomology groups $H^*(J^rY)$ of finite order jet manifolds $J^rY$.
The forthcoming Lemma \ref{3jpa} completes the proof.
\end{proof}

\begin{lem} \label{3jpa} \mar{3jpa}
De Rham cohomology of any finite-order jet manifold $J^rY$ is 
equal to that of $Y$.
\end{lem}

\begin{proof} 
Since every fibre bundle $J^rY\to J^{r-1}Y$ is affine, $J^{r-1}Y$ is
a strong deformation retract of $J^rY$, and so is $Y$ (see Appendix A).  Then,
in accordance with the Vietoris--Begle theorem \cite{bred},
cohomology $H(J^rY,\Bbb R)$ of $J^rY$ with coefficients in the
constant sheaf $\Bbb R$ coincides with that $H(Y,\Bbb R)$ of $Y$. The
well-known de Rham theorem completes the proof.
\end{proof}

Turn now to de Rham cohomology of the graded differential algebra
$\cQ^*_\infty$.  Let us consider the complex of sheaves 
\mar{lmp71} \beq
0\to \Bbb R\to
\gQ^0_\infty\op\longrightarrow^d\gQ^1_\infty\op\longrightarrow^d
\cdots
\label{lmp71}
\eeq
on $J^\infty Y$ and the de Rham complex of their structure algebras
\mar{5.13'} \beq
0\to \Bbb R\to
\cQ^0_\infty\op\longrightarrow^d\cQ^1_\infty\op\longrightarrow^d
\cdots\, .
\label{5.13'}
\eeq

\begin{prop} \label{38jp} \mar{38jp} There is an isomorphism
\be
H^*(\cQ^*_\infty)=H^*(Y) 
\ee
of de Rham cohomology $H^*(\cQ^*_\infty)$ 
of the graded differential algebra
$\cQ^*_\infty$  to that $H^*(Y)$ of the fiber bundle $Y$.
\end{prop}

\begin{proof} The complex (\ref{lmp71}) is exact due to
the Poincar\'e lemma, and is a resolution of the constant sheaf $\Bbb R$ on
$J^\infty Y$ since $\gQ^r_\infty$ are sheaves of $\cQ^0_\infty$-modules.
Then, by virtue of Theorem \ref{+132}, we have the
cohomology isomorphism  
\mar{lm80}\beq
H^*(\cQ^*_\infty)=H^*(J^\infty Y,\Bbb R). \label{lm80}
\eeq
Lemma \ref{20jpa} below completes the proof.
\end{proof}

\begin{lem} \label{20jpa} \mar{20jpa}
There is an
isomorphism 
\mar{lmp80}\beq
H^*(J^\infty Y,\Bbb R)= H^*(Y,\Bbb R)=H^*(Y) \label{lmp80}
\eeq
between cohomology $H^*(J^\infty Y,\Bbb R)$ of $J^\infty Y$ with
coefficients in the constant sheaf $\Bbb R$, that $H^*(Y,\Bbb R)$ of $Y$, and
de Rham cohomology $H^*(Y)$ of $Y$. 
\end{lem}

\begin{proof}
Since $Y$ is a strong deformation retract of $J^\infty Y$,
the first isomorphism in (\ref{lmp80}) follows from the
above-mentioned Vietoris--Begle theorem \cite{bred}, while the
second one is a consequence of the de Rham theorem.
\end{proof}

Since the graded differential algebras $\cO^*_\infty$ and $\cQ^*_\infty$
have the same de Rham cohomology, we agree
to call
\be
H^*(J^\infty Y)=H^*(\cQ^*_\infty)=H^*(\cO^*_\infty)
\ee
the de Rham cohomology 
of $J^\infty Y$.

Proposition \ref{38jp} shows that every closed form $\f\in \cQ^*_\infty$
splits into the sum
\mar{tams2}\beq
\f=\varphi +d\xi, \qquad \xi\in \cQ^*_\infty, \label{tams2} 
\eeq
where $\varphi$ is a closed form on the fiber bundle $Y$.  Accordingly,
Proposition \ref{a1} states that, if $\f$ in this splitting belongs to
$\cO^*_\infty$, so is $\xi$.
The decomposition (\ref{tams2}) will
play an important role  in the sequel.

\section{Cohomology of $d_V$}

Let us consider the vertical exact sequence of sheaves
\mar{lmp90'}\beq
0\to \gO^m_X \ar^{\pi^{\infty*}} \gQ^{0,m}_\infty \ar^{d_V}\cdots \ar^{d_V} 
\gQ^{k,m}_\infty \ar^{d_V}\cdots, \qquad 0\leq m\leq n, \label{lmp90'}
\eeq
in the variational bicomplex
(\ref{7})
and the corresponding complex of their structure algebras
\mar{lmp90}\beq
0\to \cO^m(X) \ar^{\pi^{\infty*}} \cQ^{0,m}_\infty \ar^{d_V}\cdots \ar^{d_V} 
\cQ^{k,m}_\infty \ar^{d_V}\cdots. \label{lmp90}
\eeq

\begin{prop} \label{7jp} \mar{7jp}
There is an isomorphism 
\mar{lmp01}\beq
H^*(m,d_V)=H^*(Y,\pi^*\gO^m_X) \label{lmp01}
\eeq
of cohomology
groups $H^*(m,d_V)$ of the complex (\ref{lmp90}) to cohomology groups
 $H^*(Y,\pi^*\gO^m_X)$ of $Y$ with
coefficients in the pull-back sheaf
$\pi^*\gO^m_X$ on $Y$.
\end{prop}

\begin{proof}
The exact sequence (\ref{lmp90'}) is a resolution of the pull-back sheaf
$\pi^{\infty*}\gO^m_X$ on $J^\infty Y$. Then, by virtue of Theorem \ref{+132},
we have a cohomology isomorphism
\be
H^*(m,d_V)=H^*(J^\infty Y,\pi^{\infty*}\gO^m_X). 
\ee
The isomorphism (\ref{lmp01}) follows from the facts that $Y$ is a strong
deformation retract of $J^\infty Y$ and that $\pi^{\infty*}\gO^m_X$ is the
pull-back onto $J^\infty Y$ of the sheaf $\pi^*\gO^m_X$ on $Y$ 
\cite{kash}.
\end{proof}

\begin{cor} Cohomology groups $H^{>\di Y}(m,d_V)$ vanish.
\end{cor}

The cohomology groups $H^*(m,d_V)$ have a
$C^\infty(X)$-module structure.  For instance, 
let 
\be
Y\cong X\times V\to X
\ee
be a trivial fibre bundle with a
typical fibre $V$. There is an obvious
isomorphism of $\Bbb R$-modules
\mar{10jp}\beq
H^*(m,d_V)=\gO^m_X\ot H^*(V). \label{10jp}
\eeq

\section{Cohomology of $d_H$}

Turn now to the rows of the variational bicomplex (\ref{7}). 
We have the exact sequence of sheaves 
\be
0\to \gQ^{k,0}_\infty \ar^{d_H}\gQ^{k,1}_\infty\ar^{d_H}\cdots  
\op\longrightarrow^{d_H} 
\gQ^{k,n}_\infty\ar^\tau\gE_k\to 0, \qquad k>0. 
\ee
Since the sheaves $\gQ^{k,0}_\infty$ and $\gE_k$ are fine, this is a
resolution of the fine sheaf $\gQ^{k,0}_\infty$. It states immediately the
following.

\begin{prop} \label{lmp99'} \mar{lmp99'} The cohomology
groups $H^*(k,d_H)$ of the complex   
\mar{am31}\beq
 0\to \cQ^{k,0}_\infty \ar^{d_H}\cQ^{k,1}_\infty\ar^{d_H}\cdots  
\op\longrightarrow^{d_H} 
\cQ^{k,n}_\infty\ar^{\tau} E_k\to 0, \qquad k>0, \label{am31}
\eeq
are trivial.
\end{prop}

This result at terms $\cQ^{k,<n}_\infty$ recovers that of \cite{tak2}. The
exactness of the complex (\ref{am31}) at the term $\cQ^{k,n}_\infty$ means
that, if
\be
\tau(\f)=0,
\qquad \f\in
\cQ^{k,n}_\infty,
\ee
then 
\be
\f=d_H\xi, \qquad \xi\in \cQ^{k,n-1}_\infty.
\ee
Since
$\tau$ is a projection operator, there is the $\Bbb R$-module decomposition
\mar{30jpa}\beq
\cQ^{k,n}_\infty=E_k\oplus d_H(\cQ^{k,n-1}_\infty). \label{30jpa}
\eeq

\begin{rem}
One can derive Proposition \ref{lmp99'} from Theorem \ref{+132},
without appealing to that sheaves $\gE_k$ are acyclic. 
\end{rem}

Let us consider the exact sequence of sheaves
\be
0\to \Bbb R \to \gQ^0_\infty \ar^{d_H} \gQ^{0,1}_\infty
\ar^{d_H}\cdots  
\op\longrightarrow^{d_H} 
\gQ^{0,n}_\infty
\ee
where all sheaves except $\Bbb R$ are fine. Then, from Theorem \ref{+132} and
Lemma
\ref{20jpa}, we state the following.

\begin{prop} \label{lmp99} \mar{lmp99}
Cohomology groups $H^r(d_H)$, $r<n$, of the complex
\mar{t10}\beq
0\to\Bbb R\to \cQ^0_\infty \ar^{d_H}\cQ^{0,1}_\infty\ar^{d_H}\cdots  
\op\longrightarrow^{d_H} \label{t10}
\cQ^{0,n}_\infty  
\eeq
are isomorphic to de Rham
cohomology groups $H^r(Y)$ of $Y$.
\end{prop}

This result recovers that of \cite{tak2}, but let us say something more.

\begin{prop} \label{lmp130} \mar{lmp130}
Any $d_H$-closed form $\si\in \cQ^{*,<n}_\infty$ is represented by the sum
\mar{t40}\beq
\si= h_0\varphi + d_H\xi, \qquad \xi\in \cQ^*_\infty, \label{t40}
\eeq
where $\varphi$ is a closed form on the fibre bundle
$Y$. 
\end{prop}

\begin{proof}
Due to the relation 
\mar{hn1}\beq
 h_0d=d_Hh_0, \label{hn1}
\eeq
the horizontal projection $h_0$
provides a homomorphism of the de Rham complex (\ref{5.13'}) to the
complex 
\mar{+481'}\beq
0\to\Bbb R\to \cQ^0_\infty \ar^{d_H}\cQ^{0,1}_\infty\ar^{d_H}\cdots  
\op\longrightarrow^{d_H} 
\cQ^{0,n}_\infty\ar^{d_H} 0.   \label{+481'}
\eeq
Accordingly, there is a 
homomorphism
\mar{lmp125}\beq
h_0^*: H^r(J^\infty Y) \to H^r(d_H), \qquad 0\leq r\leq n, \label{lmp125}
\eeq
of cohomology groups of these complexes. Proposition \ref{38jp} and
Proposition \ref{lmp99} show that, for $r< n$, the homomorphism (\ref{lmp125})
is an isomorphism (see the relation (\ref{1j}) below for the case
$r=n$). It follows that a horizontal form $\psi\in \cQ^{0,<n}$ is $d_H$-closed
(resp.
$d_H$-exact) if and only if $\psi=h_0\f$ where $\f$ is a closed (resp. exact)
form. The decomposition (\ref{tams2}) and Proposition
\ref{lmp99'} complete the proof.
\end{proof}

\begin{prop}
If
$\f\in\cQ^{0,<n}$ is a $d_H$-closed form, then $d_V\f=d\f$ is
necessarily $d_H$-exact.
\end{prop}

\begin{proof} 
Being nilpotent, the vertical differential $d_V$
defines a homomorphism of the complex (\ref{+481'}) to the complex
\be
 0\to \cQ^{1,0}_\infty \ar^{d_H}\cQ^{1,1}_\infty\ar^{d_H}\cdots  
\op\longrightarrow^{d_H} 
\cQ^{1,n}_\infty \ar^{d_H} 0
\ee
and, accordingly, a homomorphism of cohomology groups $H^*(d_H)\to
H^*(1,d_H)$ of these complexes. Since $H^{<n}(1,d_H)=0$, the result follows.
\end{proof}

\section{Cohomology of the variational complex}

Let us prolong the complex (\ref{t10}) to the variational complex
\mar{b317}\beq
0\to\Bbb R\to \cQ^0_\infty \ar^{d_H}\cQ^{0,1}_\infty\ar^{d_H}\cdots  
\op\longrightarrow^{d_H} 
\cQ^{0,n}_\infty  \op\longrightarrow^\dl E_1 
\op\longrightarrow^\dl 
E_2 \longrightarrow \cdots\,  \label{b317}
\eeq
of the graded differential algebra $\cQ^*_\infty$.
In accordance with Lemma \ref{lmp03}, the variational complex (\ref{tams1})
is a resolution of the constant sheaf $\Bbb R$ on $J^\infty Y$.  Then,
Theorem \ref{+132} and Lemma
\ref{20jpa} give immediately the following.

\begin{prop} \label{lmp05} \mar{lmp05}
There is an isomorphism
\mar{lmp06}\beq
H^*_{\rm var}=H^*(Y) \label{lmp06}
\eeq
between cohomology $H^*_{\rm var}$ of the variational complex (\ref{b317}) and
de Rham cohomology of the fibre bundle $Y$.
\end{prop}

The isomorphism (\ref{lmp06}) recovers the result of \cite{tak2} and that
of \cite{ander80} at terms $\cQ^{0,n}_\infty$, $E_1$, but let us say
something more.
The relation (\ref{am13}) for $\tau$ and
the relation (\ref{hn1}) for $h_0$ define  a homomorphisms of the
de Rham complex (\ref{5.13'}) of the algebra $\cQ^*_\infty$ to the variational
complex (\ref{b317}). The corresponding homomorphism of their cohomology
groups is an isomorphism. Then, in accordance with the splitting
(\ref{tams2}), we come to the following assertion which completes Proposition
\ref{lmp130}.

\begin{prop} \label{t41} \mar{t41}
Any $\dl$-closed form $\si\in\cQ^{k,n}_\infty$, $k\geq 0$, is represented by
the sum
\mar{t42}\bea
&& \si=h_0\varphi + d_H \xi, \qquad k=0, \qquad
\xi\in\cQ^{0,n-1},\label{t42a}\\ 
&& \si=\tau(\varphi) +\dl(\xi), \qquad k=1, \qquad
\xi\in \cQ^{0,n},
\label{t42b}\\
&& \si=\tau(\varphi) +\dl(\xi), \qquad k>1, \qquad
\xi\in E_{k-1},
\label{t42c}
\eea
where $\varphi$ is a closed $(n+k)$-form on $Y$. 
\end{prop}

\section{Cohomology of $\cO^*_\infty$}

Thus, we have the whole cohomology of the graded
differential algebra $\cQ^*_\infty$. The following theorem provides us with
$d_H$- and $\dl$-cohomology of the graded
differential algebra
$\cO^*_\infty$. 

\begin{theo} \label{am11} \mar{am11}
Graded differential algebra $\cO^*_\infty$  has the same $d_H$- and
$\dl$-cohomology as $\cQ^*_\infty$.
\end{theo}

\begin{proof}
Let the common symbol $D$ stand for the coboundary operators  $d_H$ and
$\dl$ of the variational bicomplex. 
Bearing in mind the decompositions (\ref{t40}), (\ref{t42a}) -- (\ref{t42c}),
it suffices to show that, if an element
$\f\in
\cO^*_\infty$ is
$D$-exact with respect to the algebra $\cQ^*_\infty$ (i.e., $\f=D\varphi$,
$\varphi\in\cQ^*_\infty$), then it is
$D$-exact in the algebra
$\cO^*_\infty$ (i.e., $\f=D\varphi'$, $\varphi'\in\cO^*_\infty$).
Lemma \ref{am12} states that,
if
$Y$ is a contractible fibre bundle and a $D$-exact form $\f$ on $J^\infty Y$
is of finite jet order
$[\f]$ (i.e., $\f\in\cO^*_\infty$), there exists an exterior form $\varphi\in
\cO^*_\infty$ on $J^\infty Y$ such that $\f=D\varphi$. Moreover, a glance at
the homotopy operators for $d_H$ and $\dl$ \cite{olver} shows that  the
jet order
$[\varphi]$ of $\varphi$ is bounded for all exterior forms $\f$ of fixed
jet order. Let us call this fact the finite exactness of the operator
$D$. Given an arbitrary fibre bundle
$Y$, the finite exactness takes place on $J^\infty Y|_U$ over any open subset
$U$ of
$Y$ which is homeomorphic to a convex open subset of $\Bbb R^{\di Y}$.
Now, we show the following.

(i) Suppose that the finite exactness of the operator $D$ takes place on
$J^\infty Y$ over open subsets
$U$, $V$ of $Y$ and their non-empty overlap $U\cap V$. Then, it is also true on
$J^\infty Y|_{U\cup V}$.

(ii) Given a family $\{U_\al\}$ of disjoint open subsets of $Y$, let us
suppose that the finite exactness takes place on $J^\infty Y|_{U_\al}$ over
every subset $U_\al$ from this family. Then, it is true on $J^\infty Y$ over
the union
$\op\cup_\al U_\al$ of these subsets.

\noindent
If the assertions (i) and (ii) hold, the finite
exactness of
$D$ on $J^\infty Y$ takes place  since one can
construct the corresponding covering of the manifold $Y$
(\cite{bred2}, Lemma 9.5). 

{\it Proof of (i)}. Let
$\f=D\varphi\in\cO^*_\infty$ be a $D$-exact form on
$J^\infty Y$. By assumption, it can be brought into the form
$D\varphi_U$ on $(\pi^\infty_0)^{-1}(U)$ and $D\varphi_V$ on
$(\pi^\infty_0)^{-1}(V)$, where
$\varphi_U$ and $\varphi_V$ are exterior forms of finite jet
order.
Due to the decompositions (\ref{t40}), (\ref{t42a}) -- (\ref{t42c}),
one can choose the forms $\varphi_U$, $\varphi_V$ such that
$\varphi-\varphi_U$ on $(\pi^\infty_0)^{-1}(U)$ and 
$\varphi-\varphi_V$ on $(\pi^\infty_0)^{-1}(V)$ are $D$-exact forms.
Let us consider their difference $\varphi_U-\varphi_V$ on 
$(\pi^\infty_0)^{-1}(U\cap V)$. It is a $D$-exact form of finite jet
order which, by assumption, can be written as 
$\varphi_U-\varphi_V=D\si$ where an exterior form
$\si$ is also of finite jet order. 
Lemma
\ref{am20} below shows that $\si=\si_U +\si_V$ where
$\si_U$ and
$\si_V$ are exterior forms of finite jet order on $(\pi^\infty_0)^{-1}(U)$ and
$(\pi^\infty_0)^{-1}(V)$, respectively. Then, putting
\be
\varphi'_U=\varphi_U-D\si_U, \qquad  
\varphi'_V=\varphi_V+ D\si_V,
\ee
we have the form $\f$  equal to
$D\varphi'_U$ on $(\pi^\infty_0)^{-1}(U)$ and
$D\varphi'_V$ on $(\pi^\infty_0)^{-1}(V)$, respectively. Since the
difference $\varphi'_U -\varphi'_V$ on $(\pi^\infty_0)^{-1}(U\cap V)$ vanishes,
we obtain $\f=D\varphi'$ on $(\pi^\infty_0)^{-1}(U\cup V)$ where 
\be
\varphi'\op=^{\rm def}\left\{
\begin{array}{ll}
\varphi'|_U=\varphi'_U, &\\
\varphi'|_V=\varphi'_V &
\end{array}\right.
\ee
is of finite jet order. 

{\it Proof of (ii)}. Let
$\f\in\cO^*_\infty$ be a $D$-exact form on
$J^\infty Y$.
The finite exactness on $(\pi^\infty_0)^{-1}(\cup
U_\al)$ holds since $\f=D\varphi_\al$ on every $(\pi^\infty_0)^{-1}(U_\al)$
and, as was mentioned above, the jet order
$[\varphi_\al]$ is bounded on the set of exterior forms
$D\varphi_\al$ of fixed jet order $[\f]$. 
\end{proof}

\begin{lem} \label{am20} \mar{am20}
Let $U$ and $V$ be open subsets of a fibre bundle $Y$ and $\si\in
\gO^*_\infty$ an exterior form of finite jet order on the non-empty overlap
$(\pi^\infty_0)^{-1}(U\cap V)\subset J^\infty Y$. Then, $\si$ splits
into  a sum $\si_U+ \si_V$ of exterior forms $\si_U$ and $\si_V$ of finite jet
order on
$(\pi^\infty_0)^{-1}(U)$ and $(\pi^\infty_0)^{-1}(V)$, respectively. 
\end{lem} 

\begin{proof}
By taking a smooth partition of unity on $U\cup V$ subordinate to the cover
$\{U,V\}$ and passing to the function with support in $V$, one gets a
smooth real function
$f$ on
$U\cup V$ which is 0 on a neighborhood of $U-V$ and 1 on a neighborhood of
$V-U$ in $U\cup V$. Let $(\pi^\infty_0)^*f$ be the pull-back of $f$ onto
$(\pi^\infty_0)^{-1}(U\cup V)$. The exterior form $((\pi^\infty_0)^*f)\si$ is
zero on a neighborhood of $(\pi^\infty_0)^{-1}(U)$ and, therefore, can be
extended by 0 to $(\pi^\infty_0)^{-1}(U)$. Let us denote it $\si_U$.
Accordingly, the exterior form
$(1-(\pi^\infty_0)^*f)\si$ has an extension $\si_V$ by 0 to 
$(\pi^\infty_0)^{-1}(V)$. Then, $\si=\si_U +\si_V$ is a desired decomposition
because $\si_U$ and $\si_V$
are of finite jet order which does not exceed that of $\si$. 
\end{proof}

It is readily observed that Theorem \ref{am11} is applied to de
Rham cohomology of $\cO^*_\infty$ whose isomorphism to that of
$\cQ^*_\infty$ has been stated by Proposition \ref{a1} and Proposition
\ref{38jp}

\section{The global inverse problem in the calculus of variations}

The variational complex (\ref{b317}) provides the algebraic approach to the
calculus of variations on fiber bundles in the class of
exterior forms of locally finite jet order \cite{bau,book,tul}. For instance,
the variational operator 
$\dl$ acting on
$\cQ^{0,n}_\infty$ is the Euler--Lagrange map, while $\dl$ acting on $E_1$ is
the Helmholtz--Sonin map. Let 
\be
L=\cL\om\in \cQ^{0,n}_\infty, \qquad \om =dx^1\w\cdots dx^n,
\ee
be a horizontal density on $J^\infty Y$. One can think of $L$ as
being a Lagrangian of locally finite order.
Then, the canonical
decomposition (\ref{30jpa}) leads to the first variational formula
\mar{+421}\beq
dL=\tau(dL) + (\id -\tau)(dL)= \dl_1(L) + d_H(\phi), \qquad \phi\in
\cQ^{1,n-1}_\infty, \label{+421}
\eeq
where 
the exterior form 
\be
\dl_1 (L)= 
(-1)^{\mid \La\mid} d_\La (\dr^\La_i\cL) \th^i\w\om 
\ee
is the Euler--Lagrange form associated with the Lagrangian $L$. 

Let us relate the cohomology isomorphism (\ref{lmp06}) to the global inverse
problem of the calculus in variations. As a particular repetition of
Proposition \ref{t41}, we come to its following solution in the class of
Lagrangians of locally finite order.

\begin{theo} \label{lmp112'} \mar{lmp112'}
A Lagrangian $L\in \cQ^{0,n}_\infty$ is variationally trivial, i.e., 
$\dl(L)=0$ if and only if 
\mar{tams3}\beq
L=h_0\varphi + d_H \xi, \qquad \xi\in \cQ^{0,n-1}_\infty, \label{tams3}
\eeq
where $\varphi$ is a closed $n$-form on $Y$ (see the expression
(\ref{t42a})). 
\end{theo}

\begin{theo} \label{lmp113'} \mar{lmp113'}
An 
Euler--Lagrange-type operator $\cE\in E_1$ satisfies the Helmholtz
condition $\dl(\cE)=0$ if and only if 
\mar{tams3'}\beq
\cE=\dl(L) + \tau(\f), \qquad L\in\cQ^{0,n}_\infty, \label{tams3'}
\eeq
where $\f$ is a closed $(n+1)$-form on $Y$ (see the expression (\ref{t42b})).
\end{theo}

Theorem \ref{lmp113'} recovers the result of \cite{ander80,tak2}.

\begin{rem}
As a consequence of Theorem \ref{lmp112'}, one obtains that 
the cohomology group
$H^n(d_H)$ of the complex (\ref{+481'}) obeys the relation
\mar{1j}\beq
H^n(d_H)/H^n(Y)=\dl(\cQ^{0,n}_\infty), \label{1j}
\eeq 
where $\dl(\cQ^{0,n}_\infty)$ is the $\Bbb R$-module of Euler--Lagrange
forms on $J^\infty Y$.
\end{rem}

Theorem \ref{am11} leads us to
the similar solution of the global inverse problem in the class of finite 
order Lagrangians. This is the case of higher order Lagrangian field theory.
Namely, the theses of Theorem \ref{lmp112'} and Theorem \ref{lmp113'} remain
true if all exterior forms in expressions (\ref{tams3}) and
(\ref{tams3'}) belong to
$\cO^*_\infty$.  Thus, the obstruction to
the exactness of the finite order calculus of variations is the same as 
for exterior forms of locally finite order, without minimizing the order
of Lagrangians. In particular, we recover the result of 
\cite{vin}.

Note that the local exactness of the calculus of
variations has been proved in the class of exterior forms of finite order by
use of homotopy operators which do not minimize the order of Lagrangians (see,
e.g., \cite{olver,tul}). The infinite variational complex of
such exterior forms on $J^\infty Y$ has been studied by many authors (see,
e.g., \cite{bau,book,olver,tul}). However, these forms on
$J^\infty Y$ fail to constitute a sheaf. Therefore, the cohomology obstruction
to the exactness of the calculus of variations has been obtained in the class
of exterior forms of locally finite jet order which make up the differential
algebra
$\cQ^*_\infty$ \cite{ander80,tak2} Several statements without
proof were announced in \cite{ander0}.
A solution of the global inverse problem in the calculus of variations in
the class of exterior forms of a fixed jet order has been suggested in
\cite{ander80} by a computation of cohomology of the fixed
order variational sequence (see  
\cite{kru90,vit} for another variant of such a variational sequence).
The key point of this computation lies in the local exactness
of the finite order variational sequence which however requires
rather sophisticated {\it ad hoc} technique in order to be reproduced 
(see also 
\cite{kru98}). Therefore, the results of  \cite{ander80} were not called into
play. The first thesis of \cite{ander80} agrees with Theorem
\ref{lmp112'} for finite order Lagrangians, but says that the
jet order of the form $\xi$ in the expression (\ref{tams3}) is $k-1$ if $L$ is
a $k$-order variationally trivial Lagrangian. The second one states that a
$2k$-order Euler--Lagrange operator can be always  associated with a $k$-order
Lagrangian.

Theorem \ref{lmp112'} and Theorem \ref{lmp113'} for elements of $\cO^*_\infty$
provide a solution of the global inverse problem in 
time-dependent mechanics treated as a particular field theory on smooth fiber
bundles over
$X=\Bbb R$ \cite{book98}. Note that, in time-dependent mechanics, the inverse
problem is more intricate than in field theory. Given a second order
dynamic equation, one studies the existence of an associated
Newtonian system and its equivalence to a Lagrangian
one \cite{book98}.  Since a fiber bundle $Y\to\Bbb R$ is trivial,  de Rham
cohomology of $Y$ is equal to that of its typical
fiber
$M$, and so is de Rham cohomology $H^*(J^\infty Y)$ of
$J^\infty Y$. 
The $d_V$-cohomology groups of the
differential algebra
$\cO^*_\infty$ are given by the isomorphism
(\ref{10jp}) such that
\be
H^*(0,d_V)=H^*(1,d_V)=C^\infty(\Bbb R)\ot H^*(M).
\ee
The
variational complex (\ref{b317}) in time-dependent mechanics takes the form
\be
0\to\Bbb R\to \cQ^0_\infty \ar^{d_t}\cQ^{0,1}_\infty
\op\longrightarrow^\dl E_1 
\op\longrightarrow^\dl 
E_2 \longrightarrow \cdots\,.
\ee 
Its cohomology coincides with de Rham
cohomology  of $M$. In particular,
Theorem \ref{lmp112'} states that a Lagrangian $L$ of time-dependent
mechanics is variationally trivial if and only if it takes the form
\be
L=(\varphi_t +\varphi_i y^i_t)dt  + d_t\xi,
\ee 
where $\varphi=\varphi_tdt + \varphi_idy^i$ is a closed 1-form on $Y$ (see
also \cite{ander0}).  

\section{Cohomology of conservation laws}

Let us concern briefly cohomology of conservation
laws in Lagrangian formalism on $J^\infty Y$, but everything below is also
true for a finite order Lagrangian formalism.  Let
$u$ be a vertical vector  field on a fibre bundle $Y\to X$, treated as a
generator of a local 1-parameter group of gauge transformations of $Y$. Its
infinite order jet prolongation 
\be
J^\infty u=d_\La u\dr_i^\La, \qquad 0\leq \nm\La,
\ee
is a derivation of the ring $\cQ^0_\infty$, and also defines the
contraction $u\rfloor\f$ and the  Lie derivative 
\be
\bL_{J^\infty u}\f\op=^{\rm def} J^\infty u\rfloor d\f + d(J^\infty u\rfloor
\f)
\ee 
of elements of the differential algebra $\cQ^*_\infty$. It is easily
justified that
\be
J^\infty u\rfloor d_H\f=- d_H(J^\infty u\rfloor\f), \qquad \f\in
\cQ^*_\infty. 
\ee

Let
$L$ be a Lagrangian on
$J^\infty Y$. By virtue of the first variational formula (\ref{+421}), the Lie
derivative of the Lagrangian $L$ along
$J^\infty u$ reads
\mar{hn7}\beq
\bL_{J^\infty u}L= J^\infty u\rfloor dL= u\rfloor \dl L - d_H (J^\infty
u\rfloor\phi),
\label{hn7}
\eeq
where 
\be
 J_u=- J^\infty u\rfloor\phi \in \cQ^{0,n-1}_\infty 
\ee
is called the symmetry current along the vector field $u$. 
If $L$ is an $r$-order Lagrangian, we come to the familiar expression for a
symmetry current
\be
J_u=-J^\infty u\rfloor\phi=h_0(J^{2r-1} u\rfloor \rho_L) + 
\varphi 
\ee
where $\rho_L$ is a $(2r-1)$-order Lepagean equivalent of the Lagrangian $L$
\cite{book,kru86}, and $\varphi$ is a $d_H$-closed form. Of course, a
symmetry current $J_u$ in the expression (\ref{hn7}) is not
defined uniquely, but up to a
$d_H$-closed form. In finite order Lagrangian formalism, one usually sets 
\be
J_u=h_0(J^{2r-1} u\rfloor \rho_L),
\ee
but the problem of a choice of a Lepagean equivalent $\rho_L$ remains
\cite{fern,book}.

If the Lie derivative (\ref{hn7})
vanishes, we obtain the weak conservation law
\be
d_HJ_u=d_\la J_u^\la\om\ap 0
\ee
on the shell Ker$\dl(L)$, i.e., the global section $d_H J_u$ of the
sheaf $\gQ^{0,n}_\infty$ on $J^\infty Y$ takes zero values at points of the
subspace Ker$\dl(L)\subset J^\infty Y$ given by the condition $\dl(L)=0$.
Then, one can say that the divergence $d_HJ_u$ is a relative
$d_H$-cocycle on the pair of topological spaces $(J^\infty Y,\Ker \dl(L))$.
Of course, it is a
$d_H$-coboundary, but not necessarily a relative
$d_H$-coboundary since $J_u\not\ap 0$. Therefore,
the divergence $d_HJ_u$ of a conserved current $J_u$ can be characterized by
elements of the relative $d_H$-cohomology group
$H^n_{\rm rel}(J^\infty Y,\Ker
\dl(L))$ of the pair $(J^\infty Y,\Ker
\dl(L))$.

For
instance, any conserved Noether current in the Yang--Mills gauge theory on a
principal bundle $P$ with a structure group $G$ is well known to reduce to a
superpotential, i.e.,
$J_u=W+d_H U$ where
$W\ap 0$
\cite{book,got92}. Its divergence $d_HJ_u$ belongs to the trivial element of
the relative cohomology group $H^n_{\rm rel}(J^2 Y,\Ker\dl(L_{\rm YM}))$,
where $Y=J^1P/G$.

Let now $N^n\subset X$ be an $n$-dimensional submanifold of $X$ with a
compact boundary $\dr N^n$. Let $s$ be a section of the fibre bundle 
$Y\to X$ and $\ol s=J^\infty s$ its infinite order jet prolongation, i.e.,
$y^i_\La\circ \ol s=d_\La s^i$, $0<|\La|$. Let us assume that $\ol s(\dr N^n)
\subset \Ker\dl(L)$. Then, the quantity
\mar{35jp}\beq
\op\int_{N^n} \ol s^*d_HJ_u= \op\int_{\dr N^n} \ol s^*J_u \label{35jp}
\eeq
depends only on the relative cohomology class of the divergence $d_HJ_u$.
For instance, in the above mentioned case of gauge theory, the quantity
(\ref{35jp}) vanishes. 

Let $N^{n-1}$ be a compact $(n-1)$-dimensional submanifold of $X$ without
boundary, and $s$ a section of $Y\to X$ such that $\ol
s(N^{n-1})\subset\Ker\dl(L)$.
Let $J_u$ and $J'_u$ be two currents in the first variational formula
(\ref{hn7}). They differ from each
other in a $d_H$-closed form $\varphi$. Then, the difference
\mar{36jp}\beq
\op\int_{N^{n-1}}\ol s^*(J_u - J'_u) \label{36jp}
\eeq
depends only on the
homology class of $N^{n-1}$ and the de Rham cohomology class of $\ol
s^*\varphi$. The latter is an image of the $d_H$-cohomology class of $\varphi$
under the morphisms
\be
H^{n-1}(d_H)\ar^{h_0} H^{n-1}(Y) \ar^{s^*} H^{n-1}(X).
\ee
In particular, if $N^{n-1}=\dr N^n$ is a boundary, the quantity (\ref{36jp})
always vanishes.

\section{Appendix A}

If $Q\to Z$ is an affine bundle coordinated by 
$(z^\la,q^i)$, the map
\be
[0,1]\times Q\ni (t,z^\la,q^i) \mapsto (z^\la, tq^i +(1-t)s^i(z)), 
\ee
where $s$ is a global section of $Q\to Z$, provides a homotopy from
$Q$ to $Z$ identified with $s(Z)\subset Q$. Similarly, 
a desired homotopy from $J^\infty Y$ to $Y$ is constructed

Let $\g_{(k)}$, $k\leq 1$, be global sections of the affine jet bundles
$J^kY\to J^{k-1}Y$. Then, we have a global section 
\mar{am4}\beq
\g: Y \ni (x^\la,y^i)\to (x^\la,y^i, y^i_\La =\g_{(|\La|)}{}^i_\La
\circ\g_{(|\La|-1)}\circ \cdots \circ \g_{(1)}) \in J^\infty Y. \label{am4}
\eeq
of the open surjection $\pi^\infty_0: J^\infty Y\to Y$.
Let us
consider the map
\mar{am5}\ben
&&[0,1]\times J^\infty Y\ni (t; x^\la, y^i, y^i_\La) \to (x^\la,
y^i,y'^i_\La)\in J^\infty Y, \qquad 0<|\La|, \label{am5}\\
&& y'^i_\La= f_k(t)y^i_\La +(1-f_k(t))\g_{(k)}{}^i_\La(x^\la,y^i,y^i_\Si),
\qquad |\Si|<k=|\La|, \nonumber
\een
where
$f_k(t)$ is a continuous monotone real function on $[0,1]$ such that
\mar{am6}\beq
f_k(t)=\left\{
\begin{array}{ll}
0, & \quad t\leq 1-2^{-k},\\
1, & \quad t\geq 1-2^{-(k+1)}.
\end{array}\right. \label{am6}
\eeq
A glance at the transition functions (\ref{55.21}) shows that, although written
in a coordinate form, this map is globally defined. It is continuous because,
given an open subset $U_k\subset J^kY$, the inverse image
of the open set
$(\pi^\infty_k)^{-1}(U_k)\subset J^\infty Y$, is the open subset
 \be
&& (t_k,1]\times (\pi^\infty_k)^{-1}(U_k) \cup
(t_{k-1},1]\times (\pi^\infty_{k-1})^{-1}(\pi^k_{k-1} [U_k\cap
\g_{(k)}(J^{k-1}Y)])\cup\cdots\\
&& \qquad \cup
[0,1]\times (\pi^\infty_0)^{-1}(\pi^k_0 [U_k\cap
\g_{(k)}\circ\cdots\circ\g_{(1)}(Y)])
\ee
of $[0,1]\times J^\infty Y$, where $[t_r,1]={\rm supp}\,f_r$. Then, the map
(\ref{am5}) is a desired homotopy from
$J^\infty Y$ to
$Y$ which is identified with its image under the global section (\ref{am4}).

\section{Appendix B}

\begin{proof} 
For $q=0$, the manifested isomorphism
follows from the fact that $H^0(Z,S)=\G(Z,S)$ for any sheaf $S$ on $Z$. To
prove other ones, let us 
replace the exact sequence 
(\ref{+131'}) with 
\be
0\to S\ar^h S_0\ar^{h^0} S_1\ar^{h^1}\cdots\ar^{h^{p-2}} S_{p-1}\ar^{h^{p-1}}
\Ker h^p\to 0
\ee
and consider the short exact sequences
\be
&& 0\to S\ar^h S_0\ar^{h^0}\Ker h^1\to 0, \\
&& 0\to \Ker h^{r-1}\ar^{\rm in} S_{r-1}\ar^{h^{r-1}}\Ker h^r\to 0, \qquad
1<r\leq p.
\ee
They give the corresponding exact
cohomology sequences 
\mar{10jpa,11}\ben
&&0\to H^0(Z,S)\to H^0(Z,S_0)\to H^0(Z,\Ker h^1)\to H^1(Z,S)\to \nonumber\\
&& \qquad H^1(Z,S_0) \to \cdots, \label{10jpa} \\
&&0\to H^0(Z,\Ker h^{r-1})\to H^0(Z,S_{r-1})\to H^0(Z,\Ker h^r)\to
\nonumber \\
&& \qquad H^1(Z,\Ker h^{r-1})\to H^1(Z, S_{r-1})\to\cdots\,.  \label{11jpa}
\een
Since sheaves $S_r$, $0\leq r< p$, are acyclic, the exact sequence
(\ref{10jpa})  falls into 
\mar{13jpa}\ben
&& 0\to H^0(Z,S)\to H^0(Z,S_0)\to H^0(Z,\Ker h^1)\to H^1(Z,S)\to 0,
\nonumber\\
&& H^k(Z,\Ker h^1)=H^{k+1}(Z,\Ker h^0), \qquad 1\leq k, \label{13jpa}
\een
and, similarly, the exact sequence (\ref{11jpa}) does
\mar{14jpa,5}\ben
&&0\to H^0(Z,\Ker h^{r-1})\to H^0(Z,S_{r-1})\to H^0(Z,\Ker h^r)\to
\nonumber \\
&& \qquad H^1(Z,\Ker h^{r-1})\to 0, \label{14jpa}\\
&& H^k(Z,\Ker h^r)=H^{k+1}(Z,\Ker h^{r-1}), \qquad 1\leq k. \label{15jpa}
\een
The equalities (\ref{15jpa}) for the couples of numbers 
$(k=m,r=q-m)$, $1\leq m\leq q-2$, 
 and the equality (\ref{13jpa})
for $k=q-1$ lead to the chain of isomorphisms 
\mar{16jpa}\beq
H^1(Z,\Ker h^{q-1})= H^2(Z,\Ker h^{q-2})=\cdots=H^q(Z,\Ker h^0)= H^q(Z,S).
\label{16jpa}
\eeq
The exact sequence (\ref{14jpa}) for $r=q$
contains the exact sequence
\mar{17jpa}\beq
H^0(Z,S_{q-1})\ar^{h_*^{q-1}} H^0(Z,\Ker h^q)\to H^1(Z,\Ker h^{q-1})\to 0.
\label{17jpa}
\eeq
Since $H^0(Z,S_{q-1})=\G(Z,S_{q-1})$ and  $H^0(Z,\Ker
h^q)=\Ker h^q_*$, the result follows from (\ref{16jpa}) and (\ref{17jpa}) for
$0<q\leq p$.
\end{proof}

\end{document}